\documentclass[onecolumn,superscriptaddress,floatfix]{revtex4}
\usepackage{graphicx,amsfonts,amssymb,amsmath,hyperref,hypcap,enumerate}
\usepackage{tikz}
\usepackage{textgreek}

\begin{document}
\title{Current-Mode Deep Level Transient Spectroscopy of a
Semiconductor Nanowire Field-Effect Transistor}
\author{Ivan Isakov}
\affiliation{London Centre for Nanotechnology, University College London, 17-19 Gordon Street, London WC1H 0AH, United Kingdom, and Department of Electronic and Electrical Engineering, University College London, London WC1E 7JE, United Kingdom}
\affiliation{Current address: Department of Physics, Blackett Laboratory, Imperial College London, London SW7 2AZ, United Kingdom}

\author{Marion J L Sourribes}
\affiliation{London Centre for Nanotechnology, University College London, 17-19 Gordon Street, London WC1H 0AH, United Kingdom, and Department of Electronic and Electrical Engineering, University College London, London WC1E 7JE, United Kingdom}

\author{Paul A Warburton}
\affiliation{London Centre for Nanotechnology, University College London, 17-19 Gordon Street, London WC1H 0AH, United Kingdom, and Department of Electronic and Electrical Engineering, University College London, London WC1E 7JE, United Kingdom}
\email{p.warburton@ucl.ac.uk}

\date{\today}





\begin{abstract}

One of the main limiting factors in the carrier mobility in semiconductor nanowires is the presence of deep trap levels. While deep-level transient spectroscopy (DLTS) has proved to be a powerful tool in analysing traps in bulk semiconductors, this technique is ineffective for the characterisation of nanowires due to their very small capacitance. Here we introduce a new technique for measuring the spectrum of deep traps in nanowires. In current-mode DLTS (“I-DLTS”) the temperature-dependence of the transient current through a nanowire field-effect transistor in response to an applied gate voltage pulse is measured. We demonstrate the applicability of I-DLTS to determine the activation energy and capture cross-sections of several deep defect states in zinc oxide nanowires. In addition to characterising deep defect states, we show that I-DLTS can be used to measure the surface barrier height in semiconductor nanowires.

\end{abstract}
\maketitle

\section{Introduction}
Semiconductor nanowires are promising candidates for application in nanoelectronics. Various nanowire-based devices including field-effect transistors \cite{Gol05, Dha11}, sensors \cite{Wan01}, axial and core-shell heterostructures \cite{Bjo02} and single-electron transistors \cite{CTh03} have been reported. The electrical properties of nanowires dramatically depend on defects located both in the nanowire core and on the nanowire surface \cite{Sourribes1}. Thus the understanding of defect behaviour is crucial for controlling nanowire transport properties. The investigation of  electrically active deep level defects in bulk semiconductors is usually carried out by means of transient techniques such as deep level transient spectroscopy (DLTS) \cite{VLa74} and its derivatives (e.g. photoinduced current spectroscopy – PICTS \cite{MartBois} and current transient spectroscopy \cite{Borsuk80}), thermostimulated current \cite{Simmon72}, electron paramagnetic resonance and others. Here the term ``deep level'' denotes a state that is located deep in the semiconductor band gap ($i.e.$ at energies lower than $E_C-k_B T$ and higher than $E_V + k_B T$, $E_C$ and $E_V$ being the conduction band minimum and valence band maximum, respectively).

Investigation of the electrically active deep defect states in nanowires by DLTS is a challenging task, mainly due to the difficulties of measuring the very small capacitances of nanowire devices. Only a few papers have therefore been published on nanowire deep levels: DLTS in combination with PICTS was used to study catalyst-related electrical defects in an array of Si nanowires \cite{Sat12}; DLTS was used to study a single GaN nanowire pn-junction \cite{Par06}; and low frequency noise spectroscopy (LFNS) was applied by Motayed {\it{et al}}. to investigate the effects of metal catalysts on Si nanowires \cite{Mot11, Sha}.

A different approach has been used to study nanoscale devices (such as thin film field-effect transistors), exploiting current-mode DLTS (I-DLTS) as opposed to the conventional capacitance-mode DLTS \cite{Lisystem02}. In I-DLTS the relaxation of the current through the channel in response to a gate voltage pulse is measured. I-DLTS is a non-destructive technique which can be carried out directly on a device in a transistor geometry and does not need any additional fabrication steps. Since nanowire field effect transistors may become constituents of future electronic devices, I-DLTS is a promising technique to study them.

In this paper we present I-DLTS measurements carried out on individual semiconductor nanowire field effect transistors. Theoretical models for conductivity relaxation driven by states on the nanowire surface are proposed. In addition to the usual characterisation of the deep traps in the semiconductor, an analytical model allows the measurement of the surface band-bending of a ZnO nanowire. These results corroborate data obtained using surface sensitive techniques such as ultraviolet photoelectron spectroscopy \cite{Che12} and surface photovoltage measurements \cite{Sou12}.

\subsection{Current-mode deep level transient spectroscopy in nanowires}
\subsubsection{I-DLTS method}

\begin{figure}[htbp]
\vspace{0cm}
\centering
 \includegraphics[width=0.75\columnwidth]{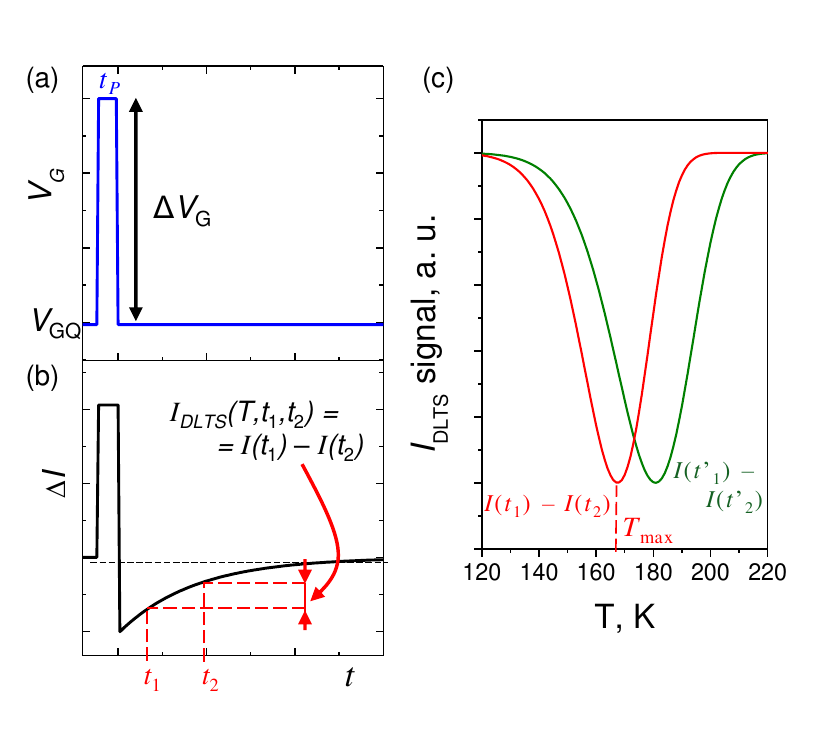}
\vspace{0cm}
\caption[I-DLTS principle of operation.]{I-DLTS principle of operation. (a) Time-dependence of the gate voltage; (b) time-dependence of the current through the nanowire; (c) I-DLTS spectra constructed using two different sets of times ($t_1$, $t_2$) and ($t'_1$, $t'_2$)}
 \label{fig0_intro}
\end{figure}

In I-DLTS measurements on nanowire FETs, a fixed drain-source voltage $V_{\text{ds}}$ is applied to the nanowire resulting in a constant current $I_{\text{ds},0}$ through the nanowire. The gate voltage is kept at a quiescent value $V_{\text{GQ}}$ and  a gate voltage pulse $\Delta V_{\text{G}}$ is applied periodically with period $T_{\text{P}}$ and width $t_{\text{P}}$ (figure~\ref{fig0_intro}.a). In n-type nanowires (like ZnO), a positive pulse $\Delta V_{\text{G}}$ populates deep trap states with electrons, which, after the end of the pulse, get emitted from the deep traps with an emission rate $e_n$, contributing to the relaxation current $\Delta I(t) = I(t) - I_{\text{ds},0}$ through the nanowire (figure~\ref{fig0_intro}.b). The I-DLTS signal is constructed by measuring the relaxation current at times $t_1$ and $t_2$ after the end of the pulse and subtracting them: $I_{\text {DLTS}} \equiv I(t_1) - I(t_2)$.

The rate of carrier emission $e_n$ from the deep traps can be expressed as: $e_n \propto \sigma_0 \, \text{exp}(-E_{\text{dl}}/k_{\text{B}}T)$, where $E_{\text{dl}}$ is the deep trap activation energy, $\sigma_0$ the trap cross-section and $k_{\text{B}}$ Boltzmann's constant. Hence the emission rate will vary with the temperature T. At sufficiently low temperatures the emission rate will be  much lower than $t_2^{-1}$, resulting in $I(t_1) - I(t_2)$ being close to zero. Conversely, at high temperatures the emission rate will be much higher than $t_1^{-1}$, also resulting in $I(t_1) - I(t_2) = 0$. At some intermediate temperature, $T_{\text{max}}$, $I_{\text {DLTS}}$ will reach a peak value (figure~\ref{fig0_intro}.c). The value of $T_{\text{max}}$ depends on the choice of $t_1$ and $t_2$. Assuming an exponential time dependence of the current $\Delta I(t) \propto \text{exp}(-e_n t)$, at temperature $T_{\text{max}}$ there is an unambiguous relation between the emission rate and the times $t_1$ and $t_2$: $e_n(T_{\text{max}})=\frac{\text{ln}(t_2/t_1)}{t_2-t_1}$ \cite{VLa74}. By choosing different values of $t_1$ and $t_2$ and measuring the temperature at which the maximum in $I_{\text{DLTS}}(T)$ occurs, we obtain the dependence of $e_n$ on temperature. From this dependence we can obtain the apparent cross-section of the trap $\sigma_0$ and its energy position in the band gap $E_{\text{dl}}$.

It will be shown later that electron emission from the electron traps generally results in a negative current transient signal as shown in figure~\ref{fig0_intro}.b. This gives rise to an I-DLTS minimum (figure~\ref{fig0_intro}.c). Conversely, hole emission will result in the opposite current transient sign, giving rise to an I-DLTS maximum. Semiconductor nanowires have different types of trap residing in the core of the nanowire, on its surface, and on the semiconductor-dielectric or semiconductor-metal interfaces. These traps will affect the current through the nanowire in different ways. An account of the physical phenomena that govern the dynamics of the carriers due to the deep surface trap states is given in the next section.

\subsubsection{Current through the nanowire FET} \label{Current_model}

In general, surface band bending affects the conductivity in a nanowire. For example, ZnO nanowires have a surface charge depletion layer due to a negative surface charge with the dominant conductivity happening in the core of the nanowire \cite{Che12, Sou12}. InAs nanowires, on the other hand, have a surface electron accumulation layer which accounts for the main contribution to the conductivity \cite{Blomers1}. 

It can be shown (Supplementary information \ref{Supplementary_Info}) that the current through a nanowire can be expressed as:

\begin{equation}
\begin{aligned}
I_{\text{ds}} =& ~ \frac{\mu_{\text{eff}} \, C_{\text{oxide}}}{L^2} [V_{\text{G}}+V_\text{T}] V_{\text{ds}};\\
V_\text{T} = & ~ Q_{\text{ss}}/ C_{\text{oxide}} + q\pi R^2 N_d L/C_{\text{oxide}},
\label{Current_main}
\end{aligned}
\end{equation}

\noindent
where $\mu_{\text{eff}}$ is the effective carrier mobility,  $V_\text{G}$ and $V_{\text{ds}}$ are the gate voltage and drain-source voltage respectively,  $V_\text{T}$
is the threshold voltage, $Q_{\text{ss}}$ the surface state charge, $N_d$ the concentration of ionised shallow donors, $L$ the distance between contacts and $R$ the radius of the nanowire. Capacitance $C_{\text {oxide}}$ is calculated based on the model of a metallic wire above a charged plane (Supplementary information \ref{Supplementary_Info}). Equation~\ref{Current_main} coincides well with the usual transistor formula in the linear regime \cite{Sze01,Gol05} with the additional term $Q_{\text{ss}}/C_{\text{oxide}}$. Because it is usually difficult to estimate the surface state concentration, this extra term can cause ambiguity in the estimation of the ionised donor concentration~$N_d$.

\begin{figure}[htbp]
\vspace{0cm}
\centering
 \includegraphics[width=0.75\columnwidth]{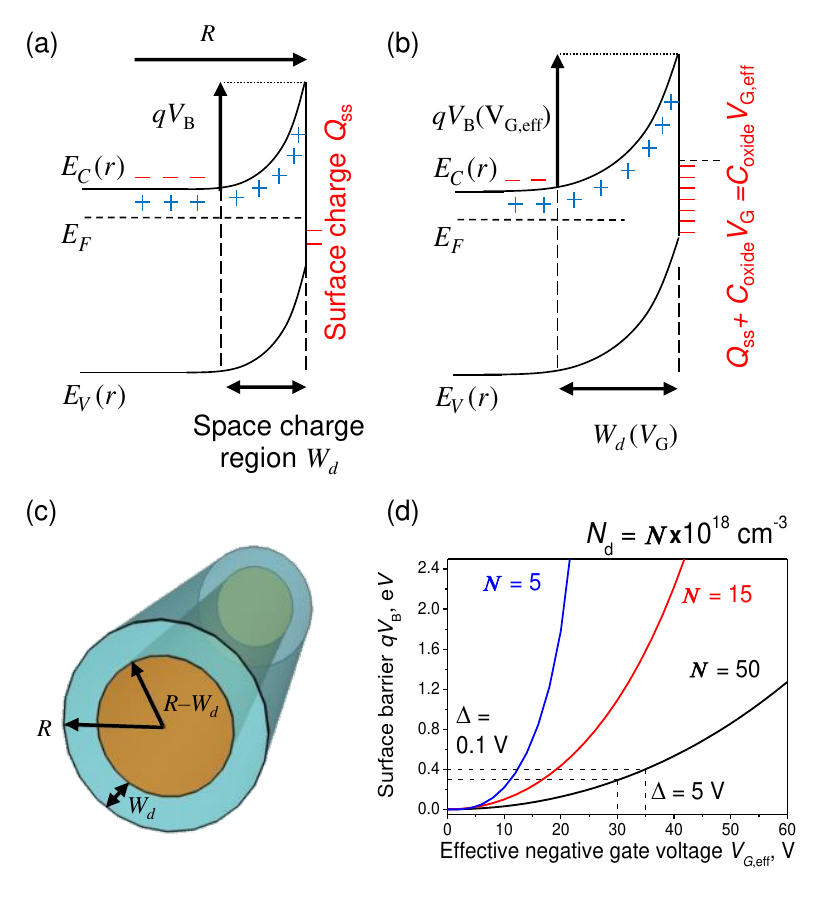}
\caption[Model of nanowire band bending for an n-type nanowire with a surface depletion region.]{Model of nanowire band bending for an n-type nanowire with a negative surface charge and surface depletion region. (a)~equilibrium condition, $V_{\text{G}}=0$; (b)~band diagram of the nanowire under applied negative gate voltage; (c)~graphic representation, with orange region being the conductive core and blue region being the depletion region; (d)~dependence of the surface barrier voltage on the effective negative gate voltage for ZnO nanowires with different donor concentrations, according to equation~\ref{CylindrCapacitance0}.}
 \label{fig1}
\end{figure}

Figure~\ref{fig1} schematically depicts a standard ZnO nanowire with a surface depletion region due to negatively charged surface states which are attributed to the absorbed oxygen molecules. Carrier capture onto the surface states depends on the surface barrier height $qV_{\text{B}} = E_b$. It can be seen in figures~\ref{fig1}.a,b that the surface barrier for electrons is affected by the gate voltage. The relationship between the gate voltage and the nanowire surface barrier height for a partially depleted nanowire can be expressed as (Supplementary information~\ref{Supplementary_Info}):

\begin{equation}
V_{\text{B}}(V_{\text{G,eff}})  =  ~  \frac{qN_d W_d(V_{\text{G,eff}}) \left[ 2R-W_d(V_{\text{G,eff}})\right] \text{ln}(R/[R-W_d(V_{\text{G,eff}})])}{ 2 \epsilon_0\epsilon_{\text {ZnO}}},
\label{CylindrCapacitance0}
\end{equation}

\noindent
where $W_d$ is the depletion region width:

\begin{equation}
W_d(V_{\text{G,eff}}) =  ~ R - \sqrt{R^2- \frac{C_{\text {oxide}} V_{\text{G,eff}}}{qN_d\pi L}}.\\
\label{CylindrCapacitance1}
\end{equation}

\noindent
Here $V_{\text{G,eff}}$ is the effective gate voltage $V_{\text{G,eff}} = Q_{\text{ss}}/C_{\text {oxide}} + V_{\text{G}}$ and $\epsilon_{\text {ZnO}}$ the dielectric constant of ZnO. Figure~\ref{fig1}.d shows the dependence of the barrier height $V_{\text{B}}$ on $V_{\text{G,eff}}$. In particular, for a typical ZnO nanowire with surface barrier of 0.3 eV and with donor concentration $N_d = 5 \cdot 10^{19}$ cm$^3$, an increase of the surface barrier by 0.1 eV ($i.e.$ from 0.3 to 0.4 V) requires the gate voltage to change by 5 V.

\subsubsection{Current transient in nanowires. Emission and capture from the surface trap} \label{Deep levels in nanowires}

Deep electron traps located in the semiconductor volume (in bulk) will affect the current and capacitance transient behaviour \cite{VLa74}. The current transients in thin film transistors are affected by both bulk and surface traps. Since nanowires exhibit very high surface-to-volume ratio, the current transient will predominantly depend on the surface states. Moreover, ZnO is known to be very surface sensitive \cite{WanQ2004}, therefore we will consider here only the surface traps. 

Let us consider surface electron traps that are uniformly distributed on the surface of the n-type nanowire (figure~\ref{fig2}). At quiescent gate voltage bias $V_{\text{GQ}}$, surface traps below the Fermi energy (with activation energy $E_{\text{ss,2}}$) are filled with electrons, while those traps above the Fermi energy (with activation energy $E_{\text{ss,1}}$) are empty (figure~\ref{fig2}.a). When a positive gate voltage pulse $\Delta V_{\text{G}}$ is applied, the Fermi energy changes its position, with surface traps below the Fermi energy capturing electrons. According to Shockley-Read-Hall statistics \cite{ShockleyRead, Hall}, the capture of electrons onto the surface states follows an exponential law with capture rate $c_n$ which depends on the surface barrier $qV_\text{B}$ (figure~\ref{fig2}.b). The change in time of the number of electrons captured on the surface state $n_{\text{ss}}(t)$ is:

\begin{equation}
\begin{aligned}
n_{\text{ss}}(t) = & ~ n_{\text{ss,max}} \, (1-\text{exp}(-c_nt));\\
c_n = & ~ \sigma_{\text{ss}} \gamma T^2 \text{exp}(-qV_\text{B}/k_{\text{B}} T),
\label{surface_states_capture}
\end{aligned}
\end{equation}

\noindent
where $\sigma_{\text{ss}}$ is the surface state capture cross-section, $n_{\text{ss,max}}$ the maximum possible number of trapped electrons and $\gamma = 2 \sqrt{3}(2\pi)^{3/2}h^{-3}k^2m^*$, with $m^*$ the effective electron mass. We assume as an approximation that the barrier height $V_{\text{B}}$ is constant in time during the charge capture. It will however depend on the gate voltage as outlined in section~\ref{Current_model}. 

At the end of the positive gate voltage pulse, the number of electrons trapped on the surface traps is $n_{\text{ss,0}} = n_{\text{ss,max}} \, (1-\text{exp}(-c_nt_p))$. When the gate voltage returns to the quiescent bias value, the filled levels start emitting electrons into the conduction band (figure~\ref{fig2}.c). The time-dependence of the number of carriers trapped on the surface levels $n_{\text{ss}}(t)$ is:

\begin{figure}[htbp]
\vspace{0cm}
 \includegraphics[width=0.75\columnwidth]{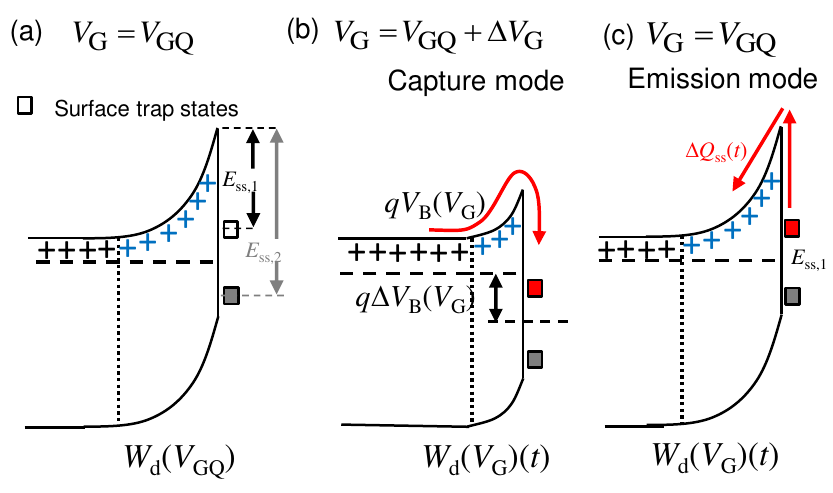}
\vspace{0cm}
\caption[Different mechanisms of current transient.]{Mechanism of the current transient in nanowires due to surface states: (a) quiescent gate voltage conditions $V_{\text{GQ}}$; (b) positive gate voltage pulse $\Delta V_{\text{G}}$ and electron capture onto surface states; (c) emission of electrons from the filled traps into the conduction band at the quiescent gate voltage.}
 \label{fig2}
\end{figure}

\begin{equation}
\begin{aligned}
n_{\text{ss}}(t) = & ~ n_{\text{ss,0}} \, \text{exp}(-e_nt);\\
e_n = & ~ \sigma_0 \gamma T^2 \,  \text{exp}(-E_{\text{ss}}/k_{\text{B}} T),
\label{carrier_transient}
\end{aligned}
\end{equation}

\noindent
where $E_{\text{ss}}$ is the energy difference between the energy level of the surface trap and the conduction band and $\sigma_0$ is an effective minority carrier capture cross section. 

The change in population of the surface levels will result in a current transient through the nanowire. Both surface and bulk traps will effectively change the surface state charge $Q_{\text{ss}}$ in equation~\ref{Current_main}. Surface charge may be represented as a sum of the constant surface charge and the transient surface charge: $Q_{\text{ss}}(t) = Q_{\text{ss,0}} + qn_{\text{ss}}(t)$. Taking into account negative charge of electrons for n-type nanowire, we get the current transient through the nanowire:

\begin{equation}
\begin{aligned}
\Delta I(t) = &-\frac{qV_{\text{ds}}\mu_{\text{eff}}}{L^2} \Delta Q_{\text{ss}}(t);\\
\Delta I_{\text{emission}}(t)= &~-\frac{\mu q}{L^2} V_{\text{ds}} n_{\text{ss,0}}\text{exp}(-e_n t);\\
\Delta I_{\text{capture}}(t)= &~\frac{\mu q}{L^2} V_{\text{ds}} n_{\text{ss,max}}\text{exp}(-c_n t).
\end{aligned}
\label{static_transient}
\end{equation}

Emission mode and capture mode \cite{Cav03} I-DLTS may be used to infer trap energies $E_{\text{ss}}$ and the barrier height $V_\text{B}$, respectively. Both methods were exploited in this work. Here, we will neglect the temperature dependences of the Fermi level position and the carrier concentration. Therefore, we assume that the values of $n_{\text{ss,0}}$ and $n_{\text{ss,max}}$ are independent of temperature for given quiescent gate voltage, gate voltage pulse magnitude and gate voltage pulse width.

\subsubsection{I-DLTS. Measurement setup}

\begin{figure}[htbp]
\centering
 \includegraphics[width=0.75\columnwidth]{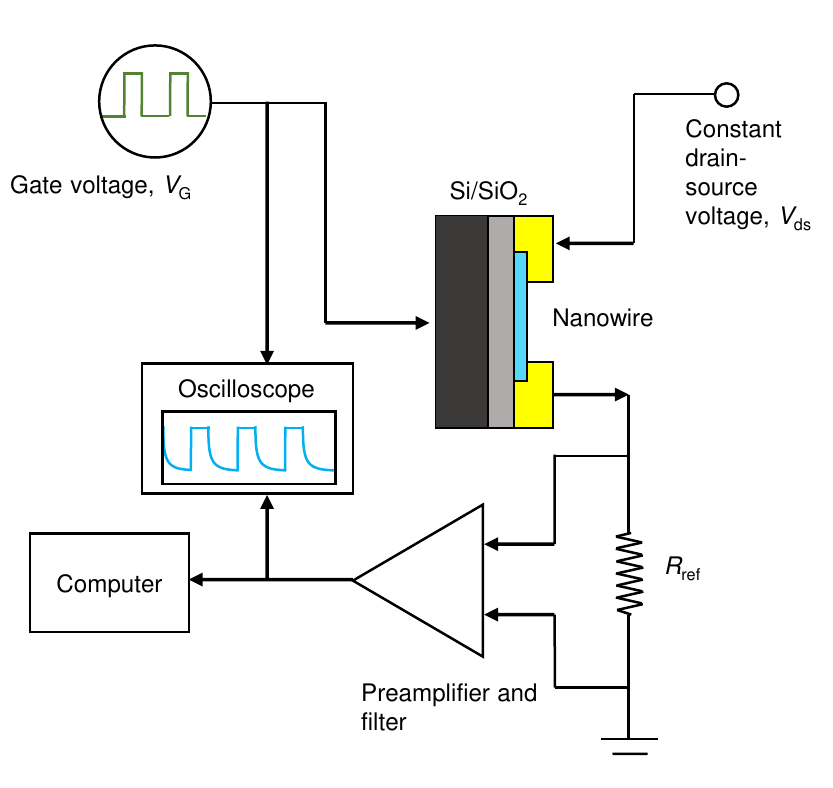}
 \caption{I-DLTS measurement setup for nanowire FET.}
 \label{fig3}
\end{figure}

The I-DLTS measurement setup is shown in figure~\ref{fig3}. A ZnO nanowire field effect transistor with back-gate is used for the measurement. The drain-source voltage through the nanowire is kept constant at $V_{\text{ds}}$ = 0.2 V. A negative quiescent gate voltage bias of $V_{\text{GQ}} = -10$ V keeps a large area of the n-type nanowire cross-section depleted.  Gate voltage pulses $\Delta V_{\text{G}}$ of 10 V amplitude and 100 $\mu$sec duration are applied to the gate of the nanowire FET with repetition rate between 0.1 and 1 kHz. The current through the nanowire is probed by the differential pre-amplifier across a reference resistor whose resistance is much smaller than the nanowire resistance. The DC component of the current is filtered out and the relaxation current is amplified. The signal is supplied to the oscilloscope, averaged and digitised. The I-DLTS signal $I_{\text{DLTS}}=I(t_1)-I(t_2)$ is measured at different temperatures.

The behaviour of the $I_{\text{DLTS}}$ peak depends on the prefactor in equation~\ref{static_transient}. It can be shown, that if we choose times $t_1$ and $t_2$ so that $t_1$ varies but the ratio $t_2/t_1$ is fixed, then the magnitude of the $I_{\text{DLTS}}/I(T)$ peak corresponding to a specific trap level will be independent of temperature.

Here for the deep level analysis, the $t_2/t_1$ ratio is fixed to 2. The time value $t_1$ ranges from 50 $\mu$sec to 4 msec.

\section{Experimental details}
\subsection{ZnO nanowire field effect transistor}
\begin{figure}[htbp]
\vspace{0cm}
\includegraphics[width=0.75\columnwidth]{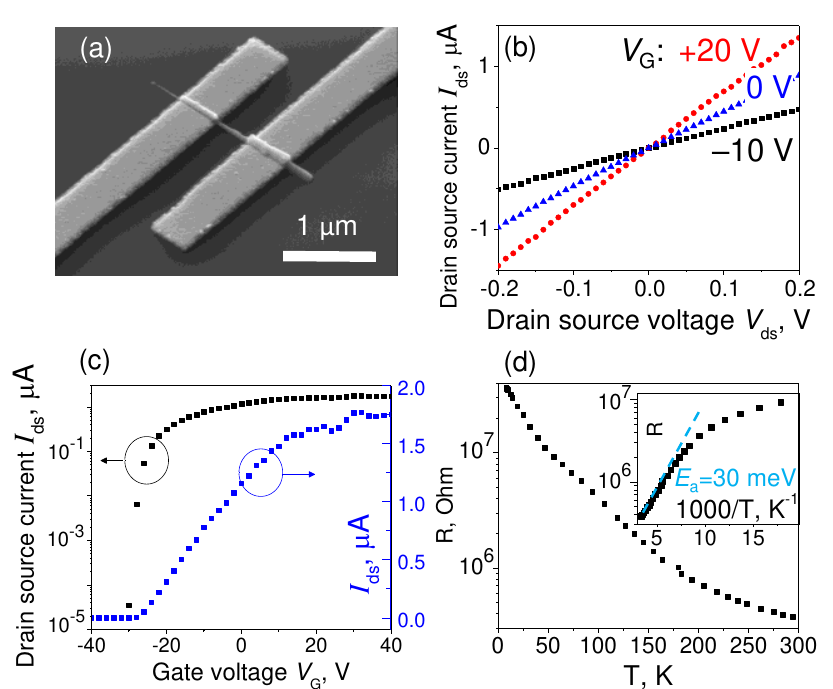}
\vspace{0cm}
\caption{(a) Scanning electron image of a connected nanowire; b) room temperature current voltage characteristics of the nanowire at different gate voltages; c) the gate voltage dependence of drain-source current; d) semi-logarithmic plot of the temperature dependence of nanowire resistance; inset: resistance on a logarithmic scale as a function of inverse temperature.}
 \label{fig4_NWFET}
\end{figure}

ZnO nanowires were grown on $\text{Al}_2 \text{O}_3(0001)$ substrates by oxygen plasma assisted molecular beam epitaxy using gold as a catalyst. Details of the growth conditions and nanowire properties are published elsewhere \cite{isa13}. Nanowires are 40--100 nm thick and 1--4 $\mu$m long. The as-grown sample was ultrasonicated in 2-propanol to remove nanowires from the substrate. A droplet of the nanowire-2-propanol solution was deposited and dried on an oxidised silicon substrate with 120 nm silicon oxide thickness. Metal contacts to the  nanowires were patterned by electron beam lithography. Nanowires were argon ion-beam milled for 30 seconds and Ti/Au contacts were sputtered onto them (figure~\ref{fig4_NWFET}.a) without breaking vacuum. The Ar ion milling was performed to remove an amorphous surface layer and thereby to decrease the contact resistance \cite{Sourribes2}. After fabricating the sample, the substrate was mounted onto the copper block of a chip carrier, which, in turn, was mounted on the end of the dip probe. Liquid helium was used to cool the sample down to 4.2~K.

Room temperature transport characteristics were measured with a Keithley 4200 semiconductor characterization system. The drain-source voltage dependence of the current of a typical ZnO nanowire is shown in figure~\ref{fig4_NWFET}.b, showing Ohmic behaviour. Four-point probe measurements showed negligible contact resistance. The resistance of the ZnO nanowires ranged from 100 k$\Omega$ to 1 M$\Omega$. The dependence of the drain source current on the gate voltage is depicted in figure~\ref{fig4_NWFET}.c. The On/Off ratio was approximately $10^6$. The field effect mobility  of the nanowire was calculated from the drain-source current using equation~\ref{Current_main}. The nanowire effective mobility at room temperature varied from 3 to 70 cm$^2$/(V$\cdot$sec) from wire to wire. The effective carrier concentration of the nanowires ranged from $10^{18}$ to $10^{19}$~cm$^{-3}$. Resistance, mobility and carrier concentration values are in agreement with data obtained by other researchers \cite{Gol05,Che12}. The temperature dependence of the 2-point probe DC resistance of a single nanowire is shown in figure~\ref{fig4_NWFET}.d. The activation energy derived from the high temperature region was 30 meV (figure~\ref{fig4_NWFET}.d, inset).

\subsection{I-DLTS. Trap characterisation}

\begin{figure}[htbp]
 \includegraphics[width=0.75\columnwidth]{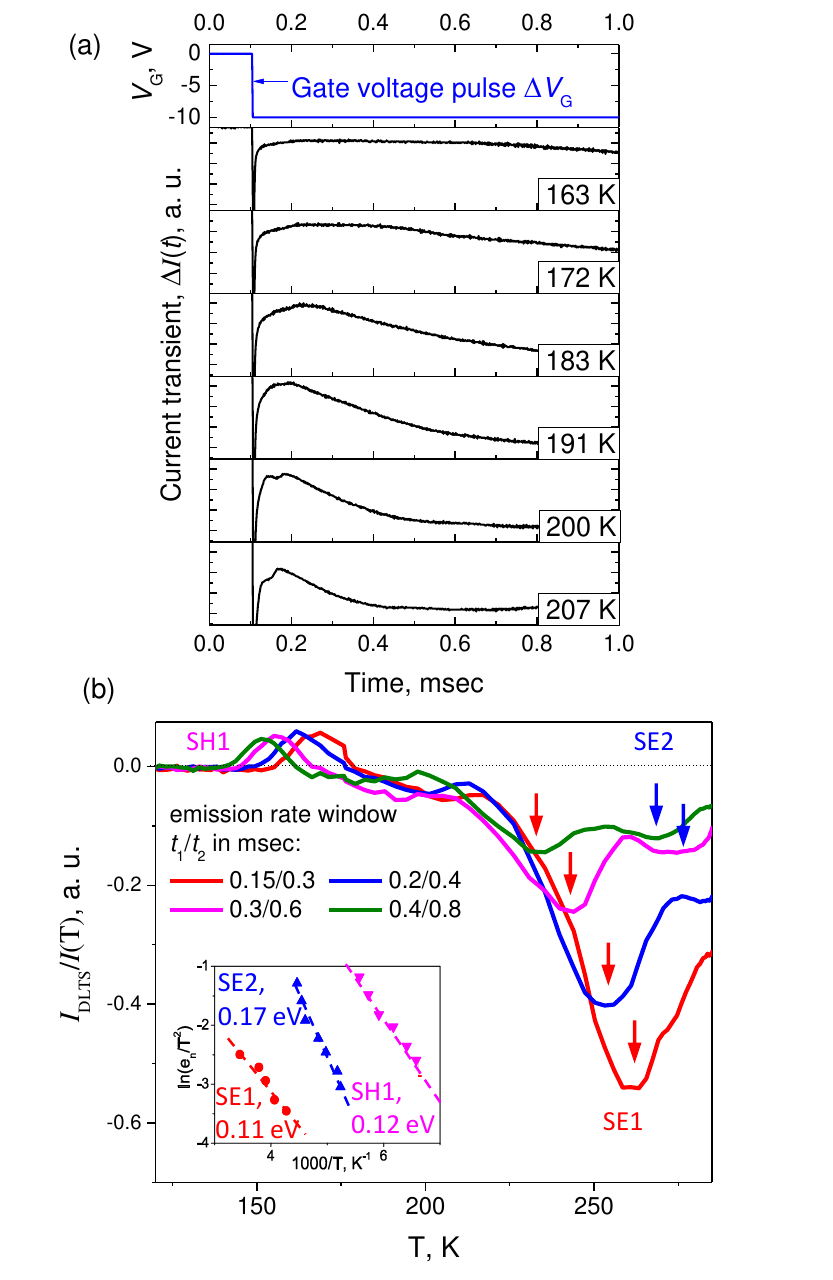}
\caption{a) Current transients of the nanowire at different temperatures for $\Delta V_{\text{G}}$ = 10~V; the negative current peak at t $<$ 0.1 msec is due to the equipment response; b) I-DLTS signal normalized to static current with deep levels indicated; inset: Arrhenius plots of levels SE1, SE2 and SH1 taken from the I-DLTS graphs in figure b).}
\label{fig6_IDLTS}
\end{figure}

I-DLTS measurements were performed on four ZnO nanowires. Current transients at different temperatures for one nanowire are presented in figure~\ref{fig6_IDLTS}.a. A transient drain current response to a gate voltage pulse similar to the one shown here was previously observed in ZnO nanowire FETs at room temperature in \cite{Mae09}. Authors attributed the current transients on a time-scale of seconds to adsorption and desorption of oxygen molecules. However, no temperature dependence was carried out. Here, the time-scales are much faster (micro- and milliseconds) and the temperatures are lower. We therefore assume that the current transients in our measurements depend only on the charging and discharging of carrier traps in the nanowire and the model for the carrier emission given above may be applied for the analysis of this data.

We shall now consider the transients in figure~\ref{fig6_IDLTS}.a. A short negative relaxation immediately after the end of the gate voltage pulse with a time constant of approximately 50 $\mu$sec is attributed to the equipment response and is similar among all the measured samples. The relaxation curves show the temperature-evolution of the current behaviour due to the deep trap. As expected from the exponential behaviour of the emission rate (equation~\ref{carrier_transient}), the emission rate increases with temperature. 


I-DLTS spectra for ZnO nanowire No.1 are shown in figure~\ref{fig6_IDLTS}.b. Peaks in this particular I-DLTS dependence may be attributed to two electron trap levels SE1 and SE2 and one hole-like trap level SH1. The activation energies and apparent capture cross sections of these traps can be determined from the Arrhenius plots (figure~\ref{fig6_IDLTS}.b, inset). Parameters of the deep levels measured on four different nanowires are listed in table~\ref{table1}. The nanowires exhibit similar hole-like traps but different electron traps. The normalised amplitude of the traps $I_{\text{DLTS}}/I(T)$ is relatively large (reaching 0.5 for the level SE1 in nanowire No.1), which indicates a high density of surface traps. It is difficult, however, to accurately estimate the density of the particular traps SE1, SE2 and SH1 due to the unknown number of equilibrium surface traps $Q_\text{ss}$ in equation~\ref{Current_main}.

\begin{table}[htbp]
\caption{Activation energies and apparent cross-section of trap levels derived from Arrhenius plots for four ZnO nanowires. For I-DLTS spectra and Arrhenius plots of all the nanowires see Supplementary Information \ref{Deep_levels_Arrh}.}
\centering
\begin{tabular}{ c  c c c }
& &  &  \\
\hline \hline
Levels & Nanowire No. & Energy, eV & Cross-section cm$^2$\\
\hline
SE1 & 1, 2 & 0.11 & 7$\times 10^{-21}$ \\
SE2 & 1, 2 & 0.15-0.2 & 4$\times 10^{-19}$--2$\times 10^{-18}$ \\
SE3* & 1 & 0.25-0.3 & (1--5)$\times 10^{-18}$ \\
E4 & 4 & 0.45 & 3$\times 10^{-12}$ \\
SH1--SH3 & 1, 2, 3, 4 & 0.10--0.20 & 2$\times 10^{-19}$--2$\times 10^{-18}$ \\
\hline
\end{tabular}

* -- Level SE3 can be seen only in spectra with emission rate windows larger than those shown in Figure~\ref{fig6_IDLTS}.b or at higher temperatures (see Supplementary Information \ref{Deep_levels_Arrh}).
\label{table1}
\end{table}

The levels observed in this work have several pronounced traits that are not expected from simple bulk deep traps. First, whereas the I-DLTS SH1 magnitude at different rate-windows stays constant at different temperatures, absolute values of the peak magnitudes attributed to the electron traps, SE1, SE2 and SE3, rapidly increase with temperature (figure~\ref{fig6_IDLTS}.b shows levels SE1 and SE2). This rapid increase in I-DLTS peak magnitude was observed in various reports and was attributed to the surface states at the un-gated area of a FET \cite{Lisystem02,ZhaoModelling1990}. Since the nanowire FETs in the present work are back-gated, the larger part of the nanowire surface is exposed to air and this affects the I-DLTS signal through the nanowire surface states. 

In addition, we can compare the measured trap signatures with trap signatures of the levels studied in bulk ZnO. ZnO films, bulk crystals and microwires have been studied by capacitance-mode DLTS with various electron traps levels observed \cite{Aur02,Que12,FSchimdt13,Que11,AHupfer14,Segh01}. Although the energies of the levels SE1, SE2 and SE3 coincide with the energies of the levels E1, E2 and E3 from these reports, the trap cross-sections are several orders of magnitude lower in the present work. Level E4, however, coincides by both energy and cross-section with the level E4 from \cite{Aur02,Que12,Segh01}, where it is attributed to oxygen vacancies. This corroborates our assumption of the surface origin of the levels SE1, SE2 and SE3.

Levels SH1, SH2 and SH3 show very similar activation energies, temperature and amplitude behaviour, and can be examined together. Hole-like levels have been observed in some n-type FETs and are usually ascribed to the surface traps, in particular to hole-traps on the un-passivated surface of the FET (\cite{Cav03} and references therein). In the case of n-type ZnO nanowires, it is very unlikely for the levels SH1 to SH3 to be real hole traps, the reasons for that being as follows. For some ZnO nanowires, the amplitude of the current transient corresponding to the hole-like level is comparable to the static current through the nanowire ($I_{\text{I-DLTS}}(T_{\text{max}})/I(T_{\text{max}}) \approx 0.1-0.5$), which, assuming the hole-like level being real hole traps, would indicate concentration of holes comparable to that of electrons. On the contrary, no inversion current was observed in the FET transfer characteristics at large negative gate biases down to --60 V, indicating a negligible concentration of holes. Therefore, the levels SH1 to SH3 cannot be real hole traps. Further study is needed to fully understand the origin of the levels SH1 to SH3.

In conclusion, all the levels observed in ZnO nanowire transistors I-DLTS, except the E4 level, appear to be surface state related. We tentatively ascribe the E4 level to the oxygen vacancies in the nanowire core. This corroborates our initial assumption about ZnO nanowires being mostly affected by the surface states. Here, we cannot unambiguously determine the exact origin of the observed surface levels. Therefore we will restrict our discussion to the phenomenological description of the trap recharging characteristics.

\subsection{I-DLTS. Surface barrier height measurement}

As it was outlined in section~\ref{Deep levels in nanowires}, capture mode I-DLTS may provide information on the nanowire surface barrier height. Any surface state that can trap electrons at positive gate voltages and de-trap electrons at negative gate voltages can be used for this purpose (equation~\ref{surface_states_capture}). The surface barrier height depends on the effective gate voltage (figure~\ref{fig2}.b.2, equation~\ref{CylindrCapacitance0}), therefore temperature of the capture-mode I-DLTS peak will depend on the quiescent gate voltage value as well.

\begin{figure}[htbp]
\centering
 \includegraphics[width=0.75\columnwidth]{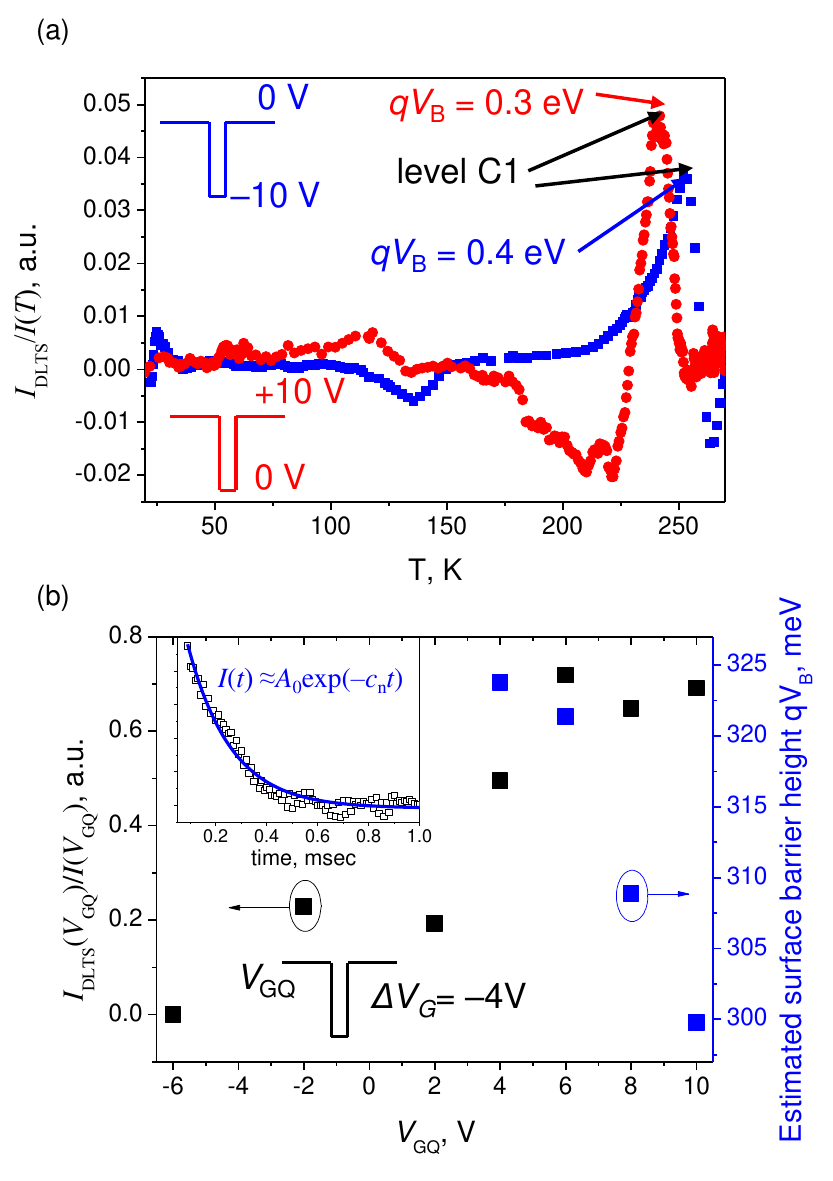}
\caption[Capture-mode I-DLTS measurement on a ZnO nanowire at different quiescent voltages.]{(a) Capture-mode I-DLTS measurement on a ZnO nanowire with different quiescent gate voltage: $V_{\text{GQ}}=+10$~V (red), $V_{\text{GQ}}=0$~V (blue), emission rate constant $t_1/t_2$ was 0.4/0.8 msec. Hole-like peaks are observed at 220--240~K; (b) quiescent voltage dependence of the capture-mode I-DLTS signal and capture activation energy, gate voltage pulse is --4~V; inset: current transient at $V_{\text{GQ}}$ = 10 V, $\Delta V_{\text{G}} = -4$~V, blue line shows the exponential fit to the experimental data.}
 \label{fig8_IDLTS_VG}
\end{figure}

Capture-mode I-DLTS was carried out on the nanowire sample No.4 (figure~\ref{fig8_IDLTS_VG}). Figure~\ref{fig8_IDLTS_VG}.a shows capture-mode I-DLTS spectra at two different quiescent voltages with the most prominent peaks at 220~K and 230~K. We assume that these two peaks correspond to the process of carrier capture over the surface barrier by the same surface trap level which we denote by C1. The surface barrier heights obtained from the I-DLTS measurements are 0.3~eV and 0.4~eV (for $V_\text{GQ}=10$~V and $V_\text{GQ}=0$~V respectively, see Supplementary Information \ref{Deep_levels_Arrh}). This values of the surface band bending in ZnO nanowire are similar to those obtained by Soudi $et\,al.$ \cite{Sou12} (0.3 eV) and Chen $et\,al.$ \cite{Che12} (0.74 eV).

According to the model in figure~\ref{fig2}.b.2, the level C1 exhibits a varying capture activation energy (and consequently, surface barrier height) which depends on the quiescent voltage. The surface barrier $qV_\text{B}$ decreases by 0.1~eV (from 0.4~eV to 0.3~eV) when the quiescent voltage increases from 0 to +10~V (figure~\ref{fig8_IDLTS_VG}.a). This value is on the same order of magnitude with that obtained from the model described in section~\ref{Current_model}, figure~\ref{fig1}.d. The capture cross-section $\sigma_{\text{ss}}$ stays approximately the same and is equal to 5$\times 10^{-18}$ cm$^{2}$.

Figure~\ref{fig8_IDLTS_VG}.b shows more detailed quiescent voltage dependence of the I-DLTS signal magnitude, with fixed gate voltage pulse of $\Delta V_{\text{G}} = -4$~V. The peak increases with quiescent voltage. Since the capture follows an exponential time-dependence (figure~\ref{fig8_IDLTS_VG}.b, inset), the capture time-constant can be inferred from the fit. It changes from 0.5 to 0.17 msec when the quiescent voltage increases from +4 to +10~V.  The capture activation energy (or barrier energy $E_b$) is calculated using equation~\ref{surface_states_capture} for $V_{\text{GQ}}>4$~V and plotted in figure~\ref{fig8_IDLTS_VG}.b, blue squares. The barrier energy monotonically decreases from 325 to 300~meV as expected from the depletion region model.

We can estimate the surface charge concentration from the barrier height  (equations~\ref{CylindrCapacitance0} and \ref{CylindrCapacitance1}). Assuming a carrier concentration of $10^{19}$ cm$^{-3}$, the surface charge density at zero gate bias is 3.5$\times 10^{12}$ cm$^{-2}$.

\section{Conclusions and prospects}

In conclusion, we have proposed and successfully implemented the I-DLTS method with gate-voltage pulsing to probe deep electronic states in individual ZnO nanowires. A variety of deep levels (SE1, SE2, SE3, E4, SH1, SH2 and SH3) were observed, with both electron-like and hole-like character. A comparison with the literature showed that levels SE1, SE2 and SE3 are most likely surface trap levels, whereas E4 is a bulk nanowire level related to oxygen vacancies. Different types of I-DLTS measurement setups (emission-mode, capture-mode and quiescent voltage dependence) were applied to the nanowire FETs. Surface barrier was established from the capture-mode I-DLTS to be 0.3--0.4 eV at 240~K.

The hole-like trap levels were discovered to affect all the nanowire FET transients. Their amplitude was found to depend on the quiescent voltage (and hence surface Fermi level). The origin of these levels is still in question.

Although the levels observed originate from the features of the ZnO nanowire FET and are most likely surface related defects, it is not clear yet whether these levels are intrinsic to ZnO nanowires, or whether they are ZnO-SiO$_2$ or ZnO-contact interface levels.

In general, I-DLTS, especially with combination with other characterization methods, has the prospects to be widely applied to nanowire research. A variety of material systems can be studied by this method: I-DLTS can be used to study the effect of the metal catalyst atom incorporation into nanowires during growth; it may give information on defect states that account for sensing properties in nanowires; and a more sophisticated analysis of the quiescent gate-voltage sweeping technique on wrap-gate FETs may give information on the spatial location of defect states in nanowires with non-uniform composition (such as axial and core-shell nanowires).

\section{Acknowledgements}
This work was funded by EPSRC grant reference EP/H005544/1. I would like to thank Dr Ed Romans and Prof Oleg Vyvenko for fruitful discussions on DLTS measurement setup and analysis of the data.

\bibliographystyle{ieeetr}
\bibliography{refs} 







\clearpage
\setcounter{figure}{0}
\newcounter{defcounter}
\setcounter{defcounter}{0}
\setcounter{section}{0}

\newenvironment{myequation}{%
\refstepcounter{defcounter}
\renewcommand\theequation{S\thedefcounter}
\begin{equation}}
{\end{equation}}

{\Large Supporting Information\par}

\section{Current through the nanowire field effect transistor} \label{Supplementary_Info}

Surface band bending determines the conductivity in a nanowire. For example, ZnO nanowires have a surface charge depletion layer due to a negative surface charge with the dominant conductivity happening in the core of the nanowire \cite{Che12, Sou12}. InAs nanowires, on the other hand, have a surface electron accumulation layer which accounts for the main contribution to the conductivity \cite{Blomers1}. A simple model for non-degenerate n-type semiconductor nanowires is given here. The effect of holes is omitted for clarity. 

We assume the nanowire to have cylindrical symmetry, a uniform distribution of ionised shallow donors with concentration $N_d$ and uniform distribution of surface states. The charge on the nanowire surface is the sum of the negative surface state charge $Q_{\text{ss}}$ and the charge induced by the gate voltage, $Q_{\text{surf}} = Q_{\text{ss}} + C_{\text {oxide}} V_{\text{G}}$. Here $V_{\text{G}}$ is the gate voltage and the oxide capacitance $C_{\text {oxide}}$ for a back-gated nanowire FET is calculated based on the model of a metallic wire above a charged plane:

\begin{myequation}
C_{\text {oxide}} = \frac{2\pi\epsilon_0\epsilon_{\text {SiO,eff}}L}{\text {ln}\big(d/R+\sqrt{(d/R)^2-1})},
\label{Ccap_0}
\end{myequation}

\noindent
where $L$ is the distance between contacts, $R$ the radius of the nanowire,  $\epsilon_{\text {SiO,eff}}$ is the effective relative permittivity of air and silicon oxide which can be taken to be equal 2.2 \cite{OWunnicke06}, and $d$ the oxide thickness. 

The charge on the outer walls of the wire should be balanced by the nanowire charge $Q_{\text{surf}} = -Q_{\text{NW}}$. The nanowire charge is constituted of the positive charge of the ionised shallow donors with concentration $N_d$ and the negative charge of the free carriers $-qL\int_0^R \! n(r)2\pi r \, \mathrm{d}r$, where $n(r)$ is the distribution of the free carrier density which depends on the conduction band profile in the nanowire. The charge in the nanowire is

\begin{myequation}
Q_{\text{NW}}=-qL\int_0^R \! n(r)2\pi r \, \mathrm{d}r + q\pi R^2N_d L.
\label{Q_NW_0}
\end{myequation}

Since $Q_{\text{surf}} = Q_{\text{ss}} + C_{\text {oxide}} V_{\text{G}}$, the charge of free carriers is equal to:

\begin{myequation}
Q_{\text{free carriers}}=qL\int_0^R \! n(r)2\pi r \, \mathrm{d}r = Q_{\text{ss}} + C_{\text {oxide}} V_{\text{G}}  + q\pi R^2 N_d L.
\label{Q_free_carriers_0}
\end{myequation}

The equation for the drain-source current $I_{\text{ds}}$ through the nanowire may be derived from \cite{Sze01}, assuming the drain-source voltage $V_{\text{ds}}$ is much smaller than the gate voltage and the surface barrier voltage:

\begin{myequation}
\begin{aligned}
I_{\text{ds}}=\frac{qV_{\text{ds}}}{L} \iint\limits_A \, \mu n\, \mathrm{d}A=\\
=\frac{qV_{\text{ds}}}{L} \int_0^R \! \mu(r) n(r) 2 \pi r\, \mathrm{d}r,
\end{aligned}
\label{Current_N_0}
\end{myequation}
 
\noindent
where $\mu(r)$ is the mobility which depends on the position (the scattering will be different at the nanowire surface and in the nanowire core) and $A$ the nanowire cross-sectional area. Next, we introduce the effective carrier mobility $\mu_{\text{eff}}$ as:

\begin{myequation}
\begin{aligned}
\mu_{\text{eff}} = \dfrac{\int_0^R \! \mu(r) n(r) 2 \pi r\, \mathrm{d}r }{\int_0^R \! n(r) 2 \pi r\, \mathrm{d}r }.
\end{aligned}
\label{Effective_mobility_0}
\end{myequation}
 
Now the current through the nanowire becomes:
\begin{myequation}
\begin{aligned}
I_{\text{ds}}=\frac{V_{\text{ds}}\mu_{\text{eff}}}{L^2} Q_{\text{free carriers}}
\end{aligned}
\label{Current_2_0}
\end{myequation}

Putting equations \ref{Q_free_carriers_0} and \ref{Current_2_0} together we get:

\begin{myequation}
\begin{aligned}
I_{\text{ds}} = \frac{V_{\text{ds}} \mu_{\text{eff}} }{L^2} \left( Q_{\text{ss}} + C_{\text{oxide}}V_{\text{G}} + q \pi R^2 N_d L \right),
\end{aligned}
\label{Qss+Ids_0}
\end{myequation}

\noindent
or, equivalently,

\begin{myequation}
I_{\text{ds}} = \frac{\mu_{\text{eff}} \, C_{\text{oxide}}}{L^2} [V_{\text{G}}+V_\text{T}] V_{\text{ds}},
\label{Current_main_0}
\end{myequation}

\noindent
where $V_\text{T} = Q_{\text{ss}}/ C_{\text{oxide}} + q\pi R^2 N_d L/C_{\text{oxide}}$
is the threshold voltage. Equation~\ref{Current_main} coincides well with the usual transistor formula in the linear regime \cite{Sze01}. The usual expression of the threshold voltage does not contain  surface charge \cite{Gol05} and is used to infer the concentration of ionised shallow donors $N_d$ in nanowires, which at high temperatures coincides with carrier concentration. The appearance of the term $Q_{\text{ss}}/C_{\text{oxide}}$ in the threshold voltage shows that the usual way of obtaining concentration of ionised shallow donors gives an incorrect result (either underestimating or overestimating the concentration depending on the sign of surface charges). The negative ($Q_{\text{ss}}<0$) or positive ($Q_{\text{ss}}>0$) surface charge will create surface depletion or accumulation layer respectively. 

Figure~\ref{fig1} schematically depicts a standard ZnO nanowire with a surface depletion region due to negatively charged surface states which are attributed to oxygen adsorption. We will consider this ZnO nanowire in order to obtain a relation between the gate voltage and the nanowire surface barrier height. 

Under applied gate bias, the surface charge becomes $Q_{\text{surf}} = Q_{\text{ss}} + C_{\text {oxide}} V_{\text{G}} = C_{\text{oxide}} V_{\text{G,eff}}$, where $V_{\text{G,eff}}$ is the effective gate voltage. The surface charge is balanced by the charge of ionised donors in the depletion region:
 
\begin{myequation}
|Q_{\text{surf}}| = Q_{\text{NW}}=qN_d\pi W_d (2R-W_d)L,
\label{SCR_charge_0}
\end{myequation}

\noindent
where $W_d$ the depletion region width. The depletion region width therefore depends on the effective gate voltage:

\begin{myequation}
W_d/R = 1 - \sqrt{1- \frac{C_{\text {oxide}} V_{\text{G,eff}}}{R^2qN_d\pi L}}.
\label{DepletionwidtheffectiveGate_0}
\end{myequation}

The charge of the depletion region may also be expressed as $Q_{\text{NW}} = C_{\text{DR}} V_{\text{B}}$, where $qV_{\text{B}} = E_b$ is the surface barrier height, $C_{\text{DR}}$ the depletion region capacitance. According to a simple cylindrical capacitor model: 

\begin{myequation}
C_{\text{DR}}  = \frac{2 \pi \epsilon_0\epsilon_{\text {ZnO}} L}{\text{ln}(R/(R-W_d))},
\label{CylindrCapacitance1_0}
\end{myequation}

\noindent
where $\epsilon_{\text {ZnO}}$ is the dielectric constant of ZnO. The barrier height can be found from combining equations~\ref{SCR_charge_0} and \ref{CylindrCapacitance1_0}:

\begin{myequation}
V_{\text{B}}  =~ \frac{qN_d W_d (2R-W_d) \text{ln}(R/(R-W_d))}{ 2 \epsilon_0\epsilon_{\text {ZnO}} },
\label{CylindrCapacitance0_0}
\end{myequation}

Combination of equations \ref{DepletionwidtheffectiveGate_0} and \ref{CylindrCapacitance0_0} gives the relationship between the nanowire surface barrier height $V_{\text{B}}$ and the effective gate voltage $V_{\text{B,eff}}$.

\section{I-DLTS spectra of all the nanowires and Arrhenius plots} \label{Deep_levels_Arrh}

\begin{figure}[!htbp]
\centering
 \includegraphics[width=0.75\columnwidth]{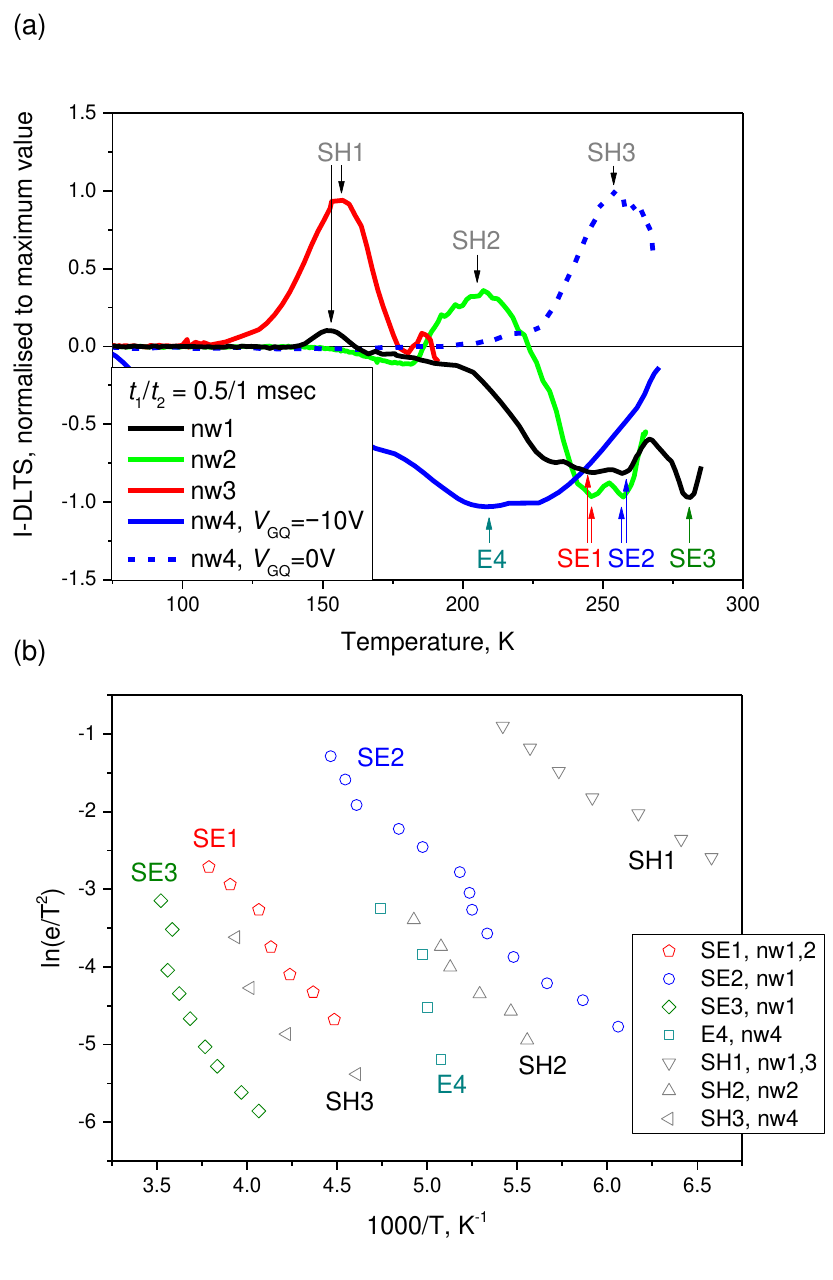}
\caption[Combined I-DLTS signal.]{(a) Combined I-DLTS spectra taken on the four nanowires with the same emission rate window. Deep electron and hole-like levels indicated. (b) Combined Arrhenius plots of all the levels measured in this work.}
 \label{figApp_IDLTS_Arrh}
\end{figure}

\begin{figure}[!htbp]
\centering
 \includegraphics[width=0.75\columnwidth]{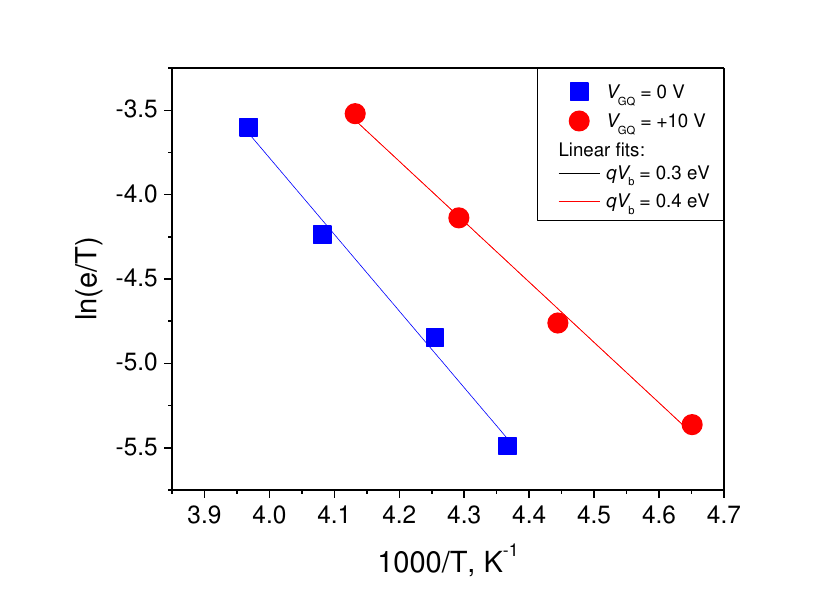}
\caption[Arrhenius on Vb measurement.]{Arrhenius plots of the peaks measured during capture mode I-DLTS for barrier height calculation.}
 \label{figApp_Arrh_Vb}
\end{figure}



\end{document}